# The Complexity of Testing Properties of Simple Games

J. Freixas[1], X. Molinero[2], M. Olsen[3], and M. Serna[4]

[1] Universitat Politècnica de Catalunya. DMA3 and EPSEM.
E-08240 Manresa, Spain. `josep.freixas@upc.edu`
[2] Universitat Politècnica de Catalunya. LSI and EPSEM.
E-08240 Manresa, Spain. `molinero@lsi.upc.edu`
[3] MADALGO[*]. Department of Computer Science. University of Aarhus.
Aabogade 34, DK 8200 Aarhus N, Denmark. `mo@madalgo.au.dk`
[4] Universitat Politècnica de Catalunya. LSI.
E-08034 Barcelona, Spain. `mjserna@lsi.upc.edu`

**Abstract.** Simple games cover voting systems in which a single alternative, such as a bill or an amendment, is pitted against the status quo. A simple game or a yes–no voting system is a set of rules that specifies exactly which collections of "yea" votes yield passage of the issue at hand. A collection of "yea" voters forms a winning coalition. We are interested on performing a complexity analysis of problems on such games depending on the game representation. We consider four natural explicit representations, winning, loosing, minimal winning, and maximal loosing. We first analyze the computational complexity of obtaining a particular representation of a simple game from a different one. We show that some cases this transformation can be done in polynomial time while the others require exponential time. The second question is classifying the complexity for testing whether a game is simple or weighted. We show that for the four types of representation both problem can be solved in polynomial time. Finally, we provide results on the complexity of testing whether a simple game or a weighted game is of a special type. In this way, we analyze strongness, properness, decisiveness and homogeneity, which are desirable properties to be fulfilled for a simple game.

## 1 Introduction

Simple game theory is a very dynamic and expanding field. Taylor and Zwicker [17] pointed out that *"few structures arise in more contexts and lend themselves to more diverse interpretations than do simple games"*. Indeed, simple games cover voting systems in which a single alternative, such as a bill or an amendment, is pitted against the status quo. In these systems, each voter responds with a vote of "yea" or "nay". A simple game or a yes–no voting system is a set of rules that specifies exactly which collections of "yea" votes yield passage of the issue at hand. A collection of "yea" voters forms a winning coalition.

---

[*] Center for Massive Data Algorithms, a Center of the Danish National Research Foundation.

Democratic societies and international organizations use a wide variety of complex rules to reach decisions. Examples, where it is not always easy to understand the consequences of the way voting is done, include the Electoral College to elect the President of the United States, the United Nations Security Council, the governance structure of the World Bank, the International Monetary Fund, the European Union Council of Ministers, the national governments of many countries, the councils in several counties, and the system to elect the major in cities or villages of many countries. Another source of examples comes from economic enterprises whose owners are shareholders of the society and divide profits or losses proportionally to the numbers of stocks they posses, but make decisions following specific rules which are voted and each shareholder has a vote proportional to the number of shares that have in the society.

One natural way to construct a simple game is to assign a (positive) real number weight to each voter, and declare a coalition to be winning precisely when its total weight meets or exceeds some predetermined quota. Simple games described in this way are said to be weighted. Yet not every simple game is weighted, every simple game can be decomposed as an intersection of some weighted games. It is related with the notion of dimension considered by Taylor and Zwicker [16,17]. The computational effort to weigh up the dimension of a game was determined by Deĭneko and Woeginger [2]: computing the dimension of a simple game is a NP-hard problem. Related work with the complexity theory in game theory appears in [14], where Prasad and Kelly provide examples of NP-completeness on determining properties of weighted majority voting games. For instance, they show that computing standard political power indices, such as absolute Banzhaf, Banzhaf–Coleman and Shapley-Shubik, are all NP-hard problems. In addition, Elkind *et al.* [3] deal with complexity analysis using representations for weighted games.

All those results relate essentially to weighted games, however there are also several alternative ways to introduce a simple game; the most natural is by giving the list of winning coalitions, then the complementary set is the set of losing coalitions and the simple game is fully described. A considerably reduction in introducing a simple game can be obtained by considering only the list of minimal winning coalitions, i.e. winning coalitions which are minimal by the inclusion operation. Coalitions containing minimal winning coalitions are also winning. Analogously, one may present a simple game by using either the set of losing coalitions or the set of maximal losing coalitions. We are interested on performing a complexity analysis depending on the game representation.

A first objective to analyze in this work is: given a particular representation of a simple game how difficult is it, from the computational viewpoint, to obtain the other three sets of coalitions? In other words, we are interested in classifying the complexity of passing from one allowable representation to another.

The second goal consists in classifying the complexity for testing whether a game is weighted assuming that the simple game is presented by one of the four sets of coalitions described above.

The third aim concerns classifying the complexity of testing whether a simple game is of a special type. Apart from weighted games there are some other subclasses of simple games which are very significant in the literature of voting systems. Strongness, properness, decisiveness and homogeneity are, among others, desirable properties to be fulfilled for a simple game.

Our results are summarized on Tables 1 and 2. Table 1 shows the complexity to pass from a given form to another one. All *explicit* forms are represented by a pair $(N, C)$ in which $N = \{1, \ldots, n\}$ for some positive integer $n$, and $C$ is the set of winning, minimal winning, losing and maximal losing coalitions. Note that it is possible to pass from winning (or losing) coalitions to minimal winning (or maximal losing) coalitions in polynomial time, but the other swaps requiere exponential time. On the other hand, given a game in a specific form, Table 2 shows the complexity on determining whether it is simple, strong, proper, weighted, homogeneous, decisive or majority. Here $(q; w)$ denotes an *integer representation* of a weighted game where $q$ is the quota and $w$ are the weights. However, there are some problems that still remain open.

| Input → <br> Output ↓ | $(N, W)$ | $(N, L)$ | $(N, W^m)$ | $(N, L^M)$ |
|---|---|---|---|---|
| $(N, W)$   | –   | EXP | EXP | EXP |
| $(N, L)$   | EXP | –   | EXP | EXP |
| $(N, W^m)$ | P   | P   | –   | EXP |
| $(N, L^M)$ | P   | P   | EXP | –   |

**Table 1.** Complexity of changing the representation form of a simple game.

| Input → | $(N, W)$ | $(N, W^m)$ | $(N, L)$ | $(N, L^M)$ | $(q; w)$ |
|---|---|---|---|---|---|
| IsSimple      | P | P      | P | P      | –      |
| IsStrong      | P | co-NPC | P | P      | co-NPC |
| IsProper      | P | P      | P | co-NPC | co-NPC |
| IsWeighted    | P | P      | P | P      | –      |
| IsHomogeneous | P | ?      | P | ?      | ?      |
| IsDecisive    | P | ?      | P | ?      | co-NPC |
| IsMajority    | P | ?      | P | ?      | co-NPC |

**Table 2.** Complexity on problems on simple games.

Up till now we deal with explicit or extensive forms (from viewpoint of giving an exhaustive list of coalitions which defines the given game), however we also consider succinct forms (a boolean formula which defines the given game).

We refer the reader to Papapdimitriou [12] for the definitions of the complexity classes P, NP, co-NP, and their subclasses of complete problems NPC and co-NPC, and the counting class #P.

## 2  Recognizing simple games

We start by giving some basic definitions on simple games (we refer the interested reader to [17] for a thorough presentation).

Simple games can be viewed as models of voting systems in which a single alternative, such as a bill or an amendment, is pitted against the status quo.

**Definition 1.** *A simple game $\Gamma$ is a pair $(N, W)$ in which $N = \{1, \ldots, n\}$ for some positive integer $n$, and $W$ is a collection of subsets of $N$ that satisfies $N \in W$, $\emptyset \notin W$, and the* monotonicity *property: $S \in W$ and $S \subseteq R \subseteq N \Rightarrow R \in W$.*

Any set of voters is called a *coalition*, the set $N$ is called the *grand coalition*, and the empty set $\emptyset$ is called the *null coalition*. Members of $N$ are called *players* or *voters*, and the subsets of $N$ that are in $W$ are called *winning coalitions*. The intuition here is that a set $S$ is a winning coalition *iff* the bill or amendment passes when the players in $S$ are precisely the ones who voted for it. A subset of $N$ that is not in $W$ is called a *losing coalition*. The collection of loosing coalitions is denoted by $L$. The set of *minimal winning coalitions* (*maximal losing coalitions*) is denoted by $W^m$ ($L^M$), where a minimal winning coalition (a maximal losing coalition) is a winning (losing) coalition all of whose proper subsets (supersets) are losing (winning). Because of monotonicity, any simple game is completely determined by its set of minimal winning coalitions. A voter $i$ is null if $i \notin S$ for all $S \in W^m$.

¿From a computational point of view a simple game can be given under different representations. In this paper we consider the following options:

– Explicit or Extensive winning form: the game is given as $(N, W)$ by providing a listing of the collection of subsets $W$.
– Explicit or Extensive minimal winning form: the game is given as $(N, W^m)$ by providing a listing of the family $W^m$. Observe that this form requires less space than the explicit winning form whenever $W \neq \{N\}$.

When we consider descriptions of a game in terms of winning coalitions in this paper, we also consider the corresponding representations for losing coalitions, replacing minimal by maximal. Thus, in addition we consider the explicit or extensive losing, and explicit or extensive maximal losing forms.

We analyse first the computational complexity of obtaining a representation of a game in a given form when a representation in another form is given. For doing so we require some additional result on families of sets.

**Lemma 1.** *Given a family of subsets $C$ of a set $N$, we can check whether it is closed under $\subseteq$ or $\supseteq$ in polynomial time.*

*Proof.* We have to check for any set $S \in C$ whether all their supersets (or subsets) are included in $C$. That is, for any $S \in C$ we have to check whether for all set $D$ such that $S \subset D \subseteq N$ (or $D \subset S$) then $D \in C$. This can be done in polynomial time in $|C|$ (see [12]).    □

**Lemma 2.** *Given a family of subsets $C$ of a set $N$, and let $\overline{C}$ be the closure of $C$ under $\subseteq$, and $\underline{C}$ be the closure of $C$ under $\supseteq$, then the families $\overline{C}^m$ and $\underline{C}^M$ can be obtained in polynomial time.*

*Proof.* Observe that, for any set $S$ in $C$ we have to check whether there is a subset (superset) of $S$ that forms part of $C$, and keep those $S$ that do not verify the property. Therefore, the complete computation can be done in polynomial time on the input length of $C$.    □

**Definition 2.** *Given a family of subsets $C$ of a set $N$, we say that it is minimal if $C = \overline{C}^m$ and maximal iff $C = \underline{C}^M$.*

As a consequence of Lemma 2 we have the following result.

**Lemma 3.** *Given a family of subsets $C$ of a set $N$, we can check whether it is maximal or minimal in polynomial time.*

Now we can state our first result for simple games given in explicit winning or losing form.

**Lemma 4.** *Given a game $\Gamma$ in explicit winning (or losing) form, the representation of $\Gamma$ in explicit minimal winning or maximal losing (maximal losing or minimal winning) form can be obtained in polynomial time.*

*Proof.* Given a game $\Gamma = (N, W)$, consider the set

$$R = \bigcup_{i=1}^{|N|} W_{-i}$$

where $W_{-i} = \{S \setminus \{i\} : i \in S \in W\}$. Observe that all the sets in $R \setminus W$ are losing coalitions, $R \setminus W \subseteq L$. We claim that $(R \setminus W)^M = L^M$. We are going to prove that in two steps:

- $(R \setminus W)^M \subseteq L^M$: Now suppose that $T \in (R \setminus W)^M$ and that $T \notin L^M$. Consequently we have that $T \in L$ and that $T \cup \{i\} \in W$ for some $i \in N$. We also have that $T \subset U$ for some $U \in L$. Due to the monotonicity we conclude that $U \cup \{i\} \in W$. This means that $U \in R \setminus W$ which contradicts that $T$ is maximal in $R \setminus W$.
- $L^M \subseteq (R \setminus W)^M$: We will show this inclusion in two steps:
  a. $L^M \subseteq R \setminus W$: If $T \in L^M$ then $T \cup \{i\} \in W$ for any $i \notin T$. Thus $T$ can be obtained from a winning coalition ($T \cup \{i\}$) from removing an element ($i$). This means that $T \in R \setminus W$ since $T$ is a losing coalition.

  b. Maximal elements in a set will also be maximal in any subset they appear in. From $L^M \subseteq R \setminus W \subseteq L$ we conclude that $L^M \subseteq (R \setminus W)^M$.

For the cost of the algorithm, observe that, given $(N, W)$, the set $R$ has cardinality at most $|N| \cdot |W|$, and thus $R$ can be obtained in polynomial time. Using Lemma 2, from $W$ and $R \setminus W$, we can compute $W^m$ and $L^M$ in polynomial time.

Analogously, when the game is given by the family of losing coalitions a symmetric argument provides the proof for explicit maximal losing or minimal winning form. □

Now we focus on simple games given in explicit minimal winning or explicit maximal losing form.

**Lemma 5.** *Given a game $\Gamma$ in explicit minimal winning (maximal losing) form, computing the representation of $\Gamma$ in explicit maximal losing (minimal winning) form requires exponential time.*

*Proof.* The following two examples show that the size of the computed family can be exponential in the size of the given one. Therefore, any algorithm that solves the problem requires exponential time.

Consider $N = \{1, \ldots, 2n\}$ and coalitions $S_i = \{2i-1, 2i\}$ for all $i = 1, \ldots, n$. Then,

(i) The simple game defined by $W^m = \bigcup_{i=1}^{n} \{S_i\}$ has

$$L^M = \{T \subseteq N : |T \cap S_i| = 1, \text{for all } i = 1, \ldots, n\}.$$

Therefore, $|W^m| = n$ and $|L^M| = 2^n$.

(ii) The simple game defined by

$$W^m = \{T \subseteq N : |T \cap S_i| = 1, \text{for all } i = 1, \ldots, n\}$$

has $L^M = \bigcup_{i=1}^{n} \{N \setminus S_i\}$. Therefore, $|W^m| = 2^n$ and $|L^M| = n$.

□

As a consequence of Lemmata 4 and 5 we have,

**Lemma 6.** *Given a game $\Gamma$ in explicit minimal winning (maximal losing) form, computing the representation of $\Gamma$ in explicit losing (winning) form requires exponential time.*

The remaining cases are again computationally hard.

**Lemma 7.** *Given a game $\Gamma$ in explicit winning (losing) form, computing the representation of $\Gamma$ in explicit losing (winning) form requires exponential time.*

*Proof.* We present two examples where the size of the computed family is exponential in the size of the given one. Let $(N, W)$ be the game, where $W = \{N\}$, then $|W| = 1$ and $|L| = 2^{|N|} - 1$. Similarly, let be the game $(N, W)$, where $L = \{\emptyset\}$, then $|W| = 2^{|N|} - 1$ and $|L| = 1$. □

Lemmatta (1)–(7) give us all results presented in Table 1. Note that most of the swaps need exponential time. Now we analyse the computational complexity of the following problems:

Name: IsSimpleE
Input: $(N, C)$
Question: Is $(N, C)$ a correct explicit representation of a simple game?

We have in total four different problems depending on the test for winning, minimal winning, losing and maximal losing. However, the problem becomes polynomial time solvable when the game representation provides an explicit description of the winning, minimal winning, losing, or maximal losing coalitions. This is a direct consequence of Lemmata 2, 1, and 3, stating that whether the family is monotonic[5] or minimal/maximal can be tested in polynomial time. This result establishes the first row of Table 2.

**Theorem 1.** *The* IsSimpleE *problem belongs to* P *for any explicit form* F: *winning, minimal winning, losing, or maximal losing.*

## 3   Problems on simple games

In this section we consider a set of decision problems related to properties that define some special types of simple games (again we refer the reader to [17]). In general we will state a property P for simple games and consider the associated decision problem which has the form:

Name: IsP
Input: A simple game $\Gamma$
Question: Does $\Gamma$ satisfy property P?

Further considerations on the complexity of such problems will be stated in terms of the input representation.

### 3.1   Recognizing strong and proper games

**Definition 3.** *A simple game $(N, W)$ is* strong *if $S \notin W$ implies $N \setminus S \in W$. A simple game that is not strong is called* weak.

Intuitively speaking, if a game is weak it has too few winning coalitions, because adding sufficiently many winning coalitions would make the game strong. Note that the addition of winning coalitions can never destroy strongness.

**Definition 4.** *A simple game $(N, W)$ is* proper *if $S \in W$ implies $N \setminus S \notin W$. A simple game that is not proper is called* improper.

---
[5] We say that a family of sets is *monotonic* iff it satisfies the monotonicity property.

An improper game has too many winning coalitions, in the sense that deleting sufficiently many winning coalitions would make the game proper. Note that the deletion of winning coalitions can never destroy properness.

When a game is both proper and strong, a coalition wins *iff* its complement loses. Therefore, in this case we have $|W| = |L| = 2^{n-1}$.

**Theorem 2.** *The* IsStrong *problem, when the input game is given in explicit losing or maximal losing form, and the* IsProper *problem, when the game is given in explicit winning or minimal winning form, can be solved in polynomial time.*

*Proof.* First observe that, given a family of subsets $F$, we can check, for any set in $F$, whether its complement is not in $F$ in polynomial time. Therefore, taking into account the definitions, we have that the IsStrong problem, when the input is given in explicit loosing form, and the IsProper problem, when the input is given in explicit winning form, are polynomial time solvable.

Thus, taking into account that

– A simple game is weak iff
$$\exists S \subseteq N : S \in L \,\wedge\, N \setminus S \in L$$
which is equivalent to
$$\exists S \subseteq N : \exists L_1, L_2 \in L^M : S \subseteq L_1 \,\wedge\, N \setminus S \subseteq L_2$$
The last assertion is equivalent to the fact that there are two maximal loosing coalitions $L_1$ and $L_2$ such that $L_1 \cup L_2 = N$.
– A simple game is *improper* iff
$$\exists S \subseteq N : S \in W \,\wedge\, N \setminus S \in W$$
which is equivalent to
$$\exists S \subseteq N : \exists W_1, W_2 \in W^m : W_1 \subseteq S \,\wedge\, W_2 \subseteq N \setminus S.$$
This last assertion is equivalent to the fact tat there are two minimal winning coalitions $W_1$ and $W_2$ such that $W_1 \cap W_2 = \emptyset$.

Observe that, given a family of subsets $F$, checking whether any one of the two conditions hold can be done in polynomial time. Thus the theorem holds also when the set of maximal loosing (or minimal winning ) coalitions is given. □

As a consequence of Lemma 4 and the previous theorem we have

**Corollary 1.** *The* IsStrong *problem, when the input game is given in explicit winning form, and the* IsProper *problem, when the game is given in explicit losing form, can be solved in polynomial time.*

Our next result states the complexity of the IsStrong problem when the game is given in the remaining forms.

**Theorem 3.** *The* IsStrong *problem is* co-NP*-complete when the input game is given in explicit minimal winning form.*

*Proof.* ¿From the definition of strong game, it is straightforward to show that the problem belongs to co-NP. We show here that the complementary problem, the IsWeak problem, when the input game is given in extensive winning form, is NP-hard. This will settle the claimed result.

We provide a polynomial time reduction from the *set splitting* problem which is known to be NP-complete [5]. An instance of the set splitting problem is a collection $C$ of subsets of a finite set $N$. The question is whether it is possible to partition $N$ into two subsets $P$ and $N \setminus P$ such that no subset in $C$ is entirely contained in either $P$ or $N \setminus P$. In other words we have to decide whether $P \subseteq N$ exists such that

$$\forall S \in C : S \nsubseteq P \wedge S \nsubseteq N \setminus P \qquad (1)$$

We transform a set splitting instance $(N, C)$ into the simple game in explicit minimal winning form $(N, C^m)$. This transformation can be computed in polynomial time according to Lemma 2. We will now show that $(N, C)$ has a set splitting iff $(N, C^m)$ is a weak game:

- Now assume that $P \subseteq N$ satisfying (1) exists. This means that $P$ and $N \setminus P$ are loosing coalitions in the game $(N, C^m)$.
- Let $P$ and $N \setminus P$ be loosing coalitions in the game $(N, C^m)$. As a consequence we have that $S \nsubseteq P$ and $S \nsubseteq N \setminus P$ for any $S \in C^m$. This implies that $S \nsubseteq P$ and $S \nsubseteq N \setminus P$ holds for any $S \in C$ since any set in $C$ contains a set in $C^m$.

□

**Theorem 4.** *The* IsProper *problem is* co-NP*-complete when the input game is given in extensive maximal losing form.*

*Proof.* ¿From Definition 4, a game is *improper* if and only if

$$\exists S \subseteq N : S \notin L \wedge N \setminus S \notin L \iff \exists S \subseteq N : \forall T_1, T_2 \in L^M : S \nsubseteq T_1 \wedge N \setminus S \nsubseteq T_2$$

Therefore the problem IsImproper belongs to NP, and the problem IsProper belongs to co-NP.

To show that the problem is also co-NP-hard we provide a reduction from the IsStrong problem for games given in extensive minimal winning form.

First observe that, if a family $C$ of subsets of $N$ is minimal then the family $\{N \setminus L : L \in C\}$ is maximal. Given a game $\Gamma = (N, W^m)$, in minimal winning form, let us consider its dual game $\Gamma' = (N, \{N \setminus L : L \in W^m\})$ given in maximal losing form. Of course $\Gamma'$ can be obtained from $\Gamma$ in polynomial time. Thus $\Gamma$ is weak iff

$$\exists S \subseteq N : S \in L(\Gamma) \wedge N \setminus S \in L(\Gamma)$$

which is equivalent to

$$\exists S \subseteq N : N \setminus S \in W(\Gamma') \wedge S \in W(\Gamma')$$

iff $\Gamma'$ is improper □

### 3.2 Recognizing weighted games

**Definition 5.** *A simple game $(N, W)$ is* weighted *if there exist a "quota" $q \in \mathbb{R}$ and a "weight function" $w : N \to \mathbb{R}$ such that each coalition $S$ is winning exactly when the sum of weights of $S$ meets or exceeds $q$.*

Note that, from the definition of simple game, we have $0 < q \leq w(N)$. Weighted games are probably the most important kind of simple games. Any specific example of a weight function $w$ and quota $q$ is said to *realize $G$* as a weighted game. A particular realization of a weighted game is denoted $(q; w_1, \ldots, w_n)$, or briefly $(q; w)$. By $w(S)$ we denote $\sum_{i \in S} w_i$.

Observe also that, from the *monotonicity property*, it is obvious that a simple game $(N, W)$ is *weighted* iff there exist a "*quota*" $q \in \mathbb{R}$ and a "*weight function*" $w : N \to \mathbb{R}$ such that

$$w(S) \geq q \quad \forall\ S \in W^m$$
$$w(S) < q \quad \forall\ S \in L^M.$$

On the other hand, although a simple game can fail to be proper and fail to be strong, this cannot happen with weighted games.

**Proposition 1.** *Any weighted game is either proper or strong.*

¿From Proposition 1, it follows that there are three kind of weighted games: proper but not strong, strong but not proper, and both strong and proper.

It is well–known that any weighted game admits an integer realization (see for instance [1]), that is, a weight function with nonnegative integer values, and a positive integer as quota. Integer realizations naturally arise; just consider the seats distributed among political parties in any voting system.

**Theorem 5.** *The IsWeighted problem can be solved in polynomial time when the input game is given in explicit winning or losing form.*

*Proof.* We provide a polynomial time reduction from the IsWeighted problem to the *Linear Programming* problem, which is known to be solvable in polynomial time [7,8].

Taking into account Lemma 2, in both cases we can obtain $W^m$ and $L^M$ in polynomial time. Once this is done we can write, again in polynomial time, the following *Linear Programming* instance $\Pi$:

```
min q
subject to   w(S) ≥ q    if  S ∈ W^m
             w(S) < q    if  S ∈ L^M
             0 ≤ w_i     for all 1 ≤ i ≤ n
             0 ≤ q
```

As $(N, W)$ is weighted iff $\Pi$ has a solution, the proposed construction is a polynomial time reduction. □

**Theorem 6.** *The* IsWeighted *problem can be solved in polynomial time when the input game is given in explicit minimal winning or maximal losing form.*

*Proof.* Given $(N, W^m)$, we are going to prove that we can decide in polynomial time whether a simple game is weighted.

For $C \subseteq N$ we let $x_C \in \{0,1\}^n$ denote the vector with the $i$'th coordinate equal to 1 if and only if $i \in C$. In polynomial time we transform $W^m$ into the boolean function $\Phi_{W^m}$ given by the DNF:

$$\Phi_{W^m}(x) = \bigvee_{S \in W^m} (\wedge_{i \in S} x_i)$$

By construction we have the following:

$$\Phi_{W^m}(x_C) = 1 \Leftrightarrow C \text{ is winning in the game given by } (N, W^m) \qquad (2)$$

Note that $\Phi_{W^m}$ is a threshold function if and only if the game given by $(N, W^m)$ is weighted:

- **only if** ($\Rightarrow$): Assume that $\Phi_{W^m}$ is a threshold function. Let $w \in \mathbb{R}^n$ be the weights and $q \in \mathbb{R}$ the threshold value. Thus we have that

  $$\Phi_{W^m}(x_C) = 1 \Leftrightarrow \langle w, x_C \rangle \geq q$$

  where $\langle \cdot, \cdot \rangle$ denotes the usual inner product. By using (2) we conclude that the game given by $(N, W^m)$ is weighted.
- **if** ($\Leftarrow$): Now assume that the game given by $(N, W^m)$ is weighted and that $(q; w)$ is a realization of such game. In this case we have the following:

  $$C \text{ is winning in the game given by } (N, W^m) \Leftrightarrow \langle w, x_C \rangle \geq q$$

  Again we use (2) and conclude that $\Phi_{W^m}$ is a threshold function.

The boolean function $\Phi_{W^m}$ is monotonic (i.e. *positive*) so according to the papers [6,13] (pages 211 and 59, respectively) we can in polynomial time decide whether $\Phi_{W^m}$ is a threshold function. Consequently we can also decide in polynomial time whether the game given by $(N, W^m)$ is weighted.

On the other hand, we can prove a similar result given $(N, L^M)$ just taking into acount that a game $\Gamma$ is weighted iff its dual game $\Gamma'$ is weighted. Then, we can use the same technique from the proof of Theorem 4. □

Note that now we can offer an alternative to the linear programming approach to decide whether the game is weighted given $(N, W)$ (first part of Theorem 5): We compute $(N, W^m)$ in polynomial time (Lemma 4) and use the procedure described in Theorem 6. However, this approach does not provide a way to determine the complexity of the IsHomogeneous problem given a game in explicit winning or losing form.

It is important to remark that it is known that "*a simple game is weighted iff it is trade robust*" [3,17]. Thus, according to Theorems 5 and 6, checking whether a simple game is trade robust can be solved in polynomial time. We might also note here that this result is different from the one obtained by Deĭneko and Woeginger [2] who proved that determinig the dimension of a simple game is a NP-hard problem.

**Corollary 2.** *The* IsTradeRobustness *problem can be solved in polynomial time when the input game is given in explicit winning, minimal winning, losing or maximal losing form.*

This subsection proves the fourth row of Table 2 (problems on polynomial).

### 3.3  Recognizing homogeneous, decisive and majority games

In this section we define the homogeneous, decisive and majority games and, afterwards, we analyse the complexity of the IsHomogeneous, IsDecisive and IsMajority problems.

**Definition 6.** *A simple game $(N, W)$ is* homogeneous *if there exist a (weighted) realization $(q; w)$ such that $q = w(S)$ for all $S \in W^m$.*

That is, a weighted game is homogeneous iff the sum of the weights of any minimal winning coalition is equal to the quota.

**Theorem 7.** *The* IsHomogeneous *problem can be solved in polynomial time when the input game is given in explicit winning or losing form.*

*Proof.* The polynomial time reduction from the IsHomogeneous problem to the *Linear Programming* problem is done in the same vein as we did to proof Theorem 5, but considering the instance $\Pi'$ obtained replacing $w(S) \geq q$, in the first set of inequalities of $\Pi$, by $w(s) = q$. It is immediate to see that the game is homogeneous iff $\Pi'$ has a solution. □

**Definition 7.** *A simple game is* decisive *(or* self–dual*, or* constant sum*) if it is proper and strong. A simple game is* indecisive *if it is not decisive.*

A related concept with decisiveness is the dualityness.

**Definition 8.** *Given a simple game $(N, W)$, its* dual game *is $(N, W^*)$, where $S \in W^*$ if and only if $N \setminus S \notin W$.*

That is, winning coalitions in the dual game are just the "blocking" coalitions in the original game. Note that $(N, W)$ is proper *iff* $(N, W^*)$ is strong, and $(N, W)$ is strong *iff* $(N, W^*)$ is proper. As a consequence, we have that a simple game $(N, W)$ is decisive *iff* $W = W^*$. On the other hand, $(N, W)$ is closed under $\subseteq$ or $\supseteq$ *iff* $(N, W^*)$ is closed under $\subseteq$ or $\supseteq$, respectively.

In the seminal work on game theory by von Neumann and Morgenstern [10] only decisive simple games were considered. Nowadays, many governmental institutions made their decisions through voting rules that are in fact decisive games. If abstention is not allowed (see [4] for voting games with abstention) ties are not possible in decisive games.

Another interesting subfamily of simple games are the so–called majority games:

**Definition 9.** *A simple game is a* majority *game if it is weighted and decisive.*

**Theorem 8.** *The* IsMajority *and the* IsDecisive *problems can be solved in polynomial time when the input game is given in explicit winning or losing form.*

*Proof.* We know that a game is both proper and strong requires that $|W| = |L| = 2^{n-1}$, and this test can be performed in polynomial time when $W$ or $L$ is given. Furthermore, under both forms, we can check whether the game is weighted in polynomial time using Theorem 5. □

Unfortunately, now we just suspect the following claims without any proof.

*Conjecture 1.* The IsDecisive problem is co-NP-complete when the input game is given in explicit minimal winning or maximal losing form.

*Conjecture 2.* The IsMajority problem is co-NP-complete when the input game is given in explicit minimal winning or maximal losing form.

We have studied the first four columns of Table 2. Henceforth we will only consider weighted games given by an integer representation $(q; w)$.

## 4   Problems on weighted games

In this section we consider weighted games which are described by an integer realization $(q; w)$. Observe that for this case majority and decisive are just the same property, so we will consider only the majority games. We analyse the complexity of problems of the type:

   Name: IsP
   Input: An integer realization $(q; w)$ of a weighted game $\Gamma$.
   Question: Does $\Gamma$ satisfy P?

We are interested in such problems associated to the properties of being strong, proper, homogeneous, and majority.

¿From now on some of the proofs are based on reductions from the NP-complete problem Partition [5], which is defined as:

   Name: Partition
   Input: $n$ integer values, $x_1, \ldots, x_n$
   Question: Is there $S \subseteq \{1, \ldots, n\}$ for which $\sum_{i \in S} x_i = \sum_{i \notin S} x_i$.

Observe that, for any instance of the PARTITION problem in which the sum of the $n$ input numbers is odd, the answer must be NO.

**Theorem 9.** *The* ISSTRONG, ISPROPER *and* ISMAJORITY *(here, equivalent to* ISDECISIVE*) problems, when the input describes a integer realization of a weighted game* $(q; w)$*, are* co-NP*-complete.*

*Proof.* ¿From the definitions of strong, proper and majority games, it is straightforward to show that the three problems belong to co-NP.

Observe that the weighted game with integer representation $(2; 1, 1, 1)$ is proper and strong, and thus decisive.

We transform an instance $x = (x_1, \ldots, x_n)$ of PARTITION problem into a realization of a weighted game according to the following schema

$$f(x) = \begin{cases} (q(x); x) & \text{when } x_1 + \cdots + x_n \text{ is even,} \\ (2; 1, 1, 1) & \text{otherwise.} \end{cases}$$

Function $f$ can be computed in polynomial time provided $q$ does, and we will use a different $q$ for each problem.

Nevertheless, independently of $q$, when $x_1 + \cdots + x_n$ is odd, $x$ is a NO input for partition, but $f(x)$ is a YES instance of ISSTRONG, ISPROPER, and ISMAJORITY, and thus a NO instance of the complementary problems.

Therefore, we have to take care only of the case in which $x_1 + \cdots + x_n$ is *even*. Assume that this is the case and let $s = (x_1 + \cdots + x_n)/2$ and $N = \{1, \ldots, n\}$. We will provide the proof that $f$ reduces PARTITION to the complementary problem.

a) ISSTRONG *problem.*

For the case of strong games, taking $q(x) = s + 1$, we have:

- If there is a $S \subset N$ such that $\sum_{i \in S} x_i = s$, then $\sum_{i \notin S} x_i = s$, thus both $S$ and $N \setminus S$ are losing coalitions and $f(x)$ is weak.
- Now assume that $S$ and $N \setminus S$ are both losing coalitions in $f(x)$ If $\sum_{i \in S} x_i < s$ then $\sum_{i \notin S} x_i \geq s + 1$, which contradicts that $N \setminus S$ is losing. Thus we have that $\sum_{i \in S} x_i = \sum_{i \notin S} x_i = s$, and there exists a partition of $x$.

Therefore, $f$ is a polytime reduction from PARTITION to ISWEAK

b) ISPROPER *problem.*

For the case of proper games we take $q(x) = s$. Then, if there is a $S \subset N$ such that $\sum_{i \in S} x_i = s$, then $\sum_{i \notin S} x_i = s$, thus both $S$ and $N \setminus S$ are winning coalitions and $f(x)$ is improper. When $f(x)$ is improper

$$\exists S \subseteq N : \sum_{i \in S} x_i \geq s \wedge \sum_{i \notin S} x_i \geq s,$$

and thus $\sum_{i \in S} x_i = s$. Thus, we have a polytime reduction from PARTITION to ISIMPROPER.

c) IsMajority *problem.*

For the case of majority games we take again $q(x) = s$. Observe that $f(x)$ cannot be weak, as in such a case there must be some $S \subseteq N$ for which,

$$\sum_{i \in S} x_i < s \land \sum_{i \notin S} x_i < s,$$

contradicting the fact that $s = (x_1 + \cdots + x_n)/2$. Therefore, the game is not majority iff it is improper, and the claim follows. □

In the same vein we have no results for the IsHomogeneous problem, but before finishing this section we introduce the following related problem:

Name: IsHomogeneousRealization
Input: An integer realization $(q; w)$ of a weighted game $\Gamma$.
Question: Is $(q; w)$ an homogeneous realization?

**Theorem 10.** *The* IsHomogeneousRealization *problem can be solved in polynomial time.*

*Proof.* Given the weights $w$, Rosenmüller [15] presents an algorithm that computes all $q$ such that $(q; w)$ is a homogeneous realization. The analysis on the complexity of the algorithm is omitted.

It is not hard to construct a dynamic programming algorithm – based on Lemma 1.1 (also called *the Basic Lemma*) in [15] – that runs in polynomial time and checks whether the integer realization $(q; w)$ is a homogeneous realization.
□

Note that, given an integer realization $(q; w)$ of a weighted game, we have not proven that this game is homogeneous, we have just proven whether such realization is homogeneous. The IsHomogeneous problem still remains open.

## 5    Succinct representations

Besides extensive representations there are other representations to present simple games. For instance, using a Boolean function, there is the so-called succinct representation [9].

¿From a computational point of view, a simple game can also be given under a succinct representation:

– Succinct winning form: the game is given by $(N, \Phi)$ where $\Phi$ is a boolean formula on $|N|$ variables providing a compact description of the sets in $W$. We have
$$\Phi : \{0, 1\}^{|N|} \to \{0, 1\}$$
such that
$$\Phi(x_1, \ldots, x_n) = 1 \iff \exists S \in W; S = \bigcup_{x_i = 1} \{i\}.$$

  In this way $W$ is identified with the truth assignments that satisfy the formula $\Phi$.
– Succinct minimal winning form: the game is given by $(N, \Phi)$ but now $\Phi$ describes the family $W^m$. Observe again that this form might require less space than the previous one whenever $W \neq \{N\}$.

The hardness of the problems associated to the succinct form came from reductions from problems on boolean formula.

**Theorem 11.** *The* ISSIMPLES-F *problem is* co-NP *complete for any succinct form* F: *winning, minimal winning, losing, or maximal losing.*

*Proof.* Observe that a set $W(L)$ is not monotonic iff there are two sets $S_1$ and $S_2$ such that $S_1 \subseteq S_2$ but $S_1 \in W$ and $S_2 \notin W$ ($S_1 \notin L$ and $S_2 \in L$). When the game is given in succinct winning or losing form, these tests can be done by guessing two truth assignments $x_1$ and $x_2$ and checking that $x_1 < x_2$, $\Phi_W(x_1) = 1$ and $\Phi_W(x_2) = 0$ ($\Phi_L(x_1) = 0$ and $\Phi_W(x_2) = 1$). Both properties can be checked in polynomial time once $S_1$ and $S_2$ are given.

In the case that $\Phi$ represents $W^m (L^M)$ we have to check that the represented set is minimal (maximal). Observe that $\Phi$ does not represent a minimal (maximal) set if there are $\alpha, \beta \in \{0,1\}^n$ with $\alpha < \beta$ such that $\Phi(\beta) = 1$ and $\Phi(\alpha) = 1$.

Therefore, all the problems of recognizing succinct forms belong to co-NP.

The *canonical order* on the set of truth assignments of a set of $n$ variables is defined as follows: $x \leq y$ iff $x_i \leq y_i$ for all $i$; and, $x < y$ iff $x \leq y$ and $x_i \neq y_i$ for some $i$. A boolean formula is *monotonic* if for any pair of truth assignments $x, y$, such that $x < y$ in canonical order, we have that $\Phi(x) < \Phi(y)$ (assuming that false $<$ true). The later problem is co-NP-complete (even for DNF, disjunctive normal form, formulas) [9]. Observe that, if $\Phi_W$ is a boolean formula representing $W$, then we have that $(N,W)$ is simple iff $\Phi_W$ is monotonic. Furthermore, if $\Phi_L$ is a boolean formula representing $L$, then we have that $(N, L)$ is simple iff $\neg \Phi_L$ is monotonic. Thus both associated problems are co-NP-hard.

Recall that the SAT problem asks whether a given boolean formula has a satisfying assignment. SAT is a well known NP-complete problem. We consider here the variation of SAT in which the input is a boolean formula $\Phi$ such that $\Phi(0^n) = 0$, of course this variation is also NP-complete. Given a boolean formula $\Phi$ on $n$ variables with $\Phi(0^n) = 0$, we construct a new formula on $n+1$ variables defined as follows:

$$\Psi(\alpha a) = \begin{cases} \Phi(\alpha) & \text{if } a = 1 \\ 0 & \text{if } a = 0 \text{ and } |\alpha|_1 > 1 \\ 0 & \text{if } a = 1 \text{ and } |\alpha|_1 = 1 \end{cases}$$

Observe that $\Phi$ has a satisfying assignment iff $\Psi$ represents a non minimal set (or a non maximal set). Thus the remaining problems are co-NP-hard.  □

## 6   Conclusions

Given a simple game, we have studied the complexity to pass from an explicit form to another one. All explicit forms that we have considered are represented by a pair $(N, C)$ in which $N = \{1, \ldots, n\}$ for some positive integer $n$, and $C$ is the set of winning, minimal winning, losing and maximal losing coalitions.

Given a simple game in an explicit form $(N, C)$, we have studied the complexity to decide whether it is strong, proper, weigthed, homogeneous, decisive or majority. In the same vein, given a weighted game in an integer representation $(q; w)$, we have also considered the complexity to decide whether it is strong, proper, homogeneous or majority (here to be majority is equivalent to be decisive).

There are some interesting open problems in which we are working on. For instance, given a game in explicit minimal winning or maximal losing form, we conjecture that to decide whether it is a decisive (or a majority) game is co-NP-complete.

We would also like to remark that our study can be enlarged by considering new forms to present a simple game. For example, blocking coalitions and minimal blocking coalitions provide an alternative way to fully describe a simple game. Precisely, a blocking coalition wins whenever its complementary loses. Of course, there are other presentations for a simple game, which can be implemented using a Boolean function. Probably the most natural way to do so is by means of the multilinear extension of a simple game [11]. However, we leave this part for future research.

## Acknowledgements

Josep Freixas was partially supported by Grant MTM 2006–06064 of "Ministerio de Ciencia y Tecnología y el Fondo Europeo de Desarrollo Regional" and SGRC 2005-00651 of "Generalitat de Catalunya".

Xavier Molinero was partially supported by project TIN2005-05446 (ALINEX) of "Ministerio de Educacion y Ciencia " and SGRC 2005-00516 of "Generalitat de Catalunya".

Maria Serna was partially supported by by FET pro-active Integrated Project 15964 (AEOLUS), the spanish projects MEC TIN2005-09198-C02-02 (ASCE) and TIN2005-25859-E, and SGRC 2005-00516 of "Generalitat de Catalunya".

We would also like to aknowledge Gerth S. Brodal from University of Aarhus for valuable comments and constructive criticism.